\documentclass[runningheads]{llncs}

\usepackage{epsf,graphicx,amsfonts,amsmath,amssymb}
\usepackage{soul}
\usepackage[latin1]{inputenc}
\usepackage[draft]{changes}
\usepackage[draft,textsize=tiny]{todonotes}

\font\bbfnt=msbm10

\def\bbR{\mbox{\bbfnt R}}

\newcommand{\mb}[1]{\mbox{\bfseries \itshape #1}}

\newcommand{\1}{\Bar{1}}
\newcommand{\2}{\Bar{\Bar{1}}}

\newtheorem{coro}{Corollary}

\newtheorem{prop}{Proposition}

\begin{document}

\title{Observability of dynamical networks from graphic and
symbolic approaches}

\titlerunning{Observability of dynamical networks}

\author{Irene Sendi\~na-Nadal\inst{1,2}\orcidID{0000--0003-0432-235X},  \\
\and 
Christophe Letellier\inst{3}\orcidID{0000--0003-3603-394X}} 

\authorrunning{I. Sendi\~na-Nadal and C. Letellier} 

\institute{Complex Systems Group \& GISC, Universidad Rey Juan Carlos,
28933 M\'ostoles, Madrid, Spain
\email{irene.sendina@urjc.es}
\and
Center for Biomedical Technology, Universidad Polit\'ecnica de Madrid, 28223 
Pozuelo de Alarc\'on, Madrid, Spain
\and
Normandie Universit\'e CORIA, Campus Universitaire du Madrillet, 
F-76800 Saint-Etienne du Rouvray, France 
\email{christophe.letellier@coria.fr}
}

\maketitle

\begin{abstract}
A dynamical network, a graph whose nodes are dynamical systems,
is usually characterized by a large dimensional space which is not
always accesible due to the impossibility of measuring all the variables spanning the state
space. Therefore, it is of the utmost importance to determine a reduced set of
variables providing all the required information for 
non-ambiguously distinguish its different states. Inherited from control 
theory, one possible approach is based on the use of the observability 
matrix defined as the Jacobian matrix of the change of coordinates between 
the original state space and the space reconstructed from the measured 
variables. The observability of a given system can be accurately 
assessed by symbolically computing the complexity of the determinant of the 
observability matrix and quantified by symbolic observability coefficients.
In this work, we extend the symbolic observability, previously developed for 
dynamical systems, to networks made of coupled $d$-dimensional node dynamics 
($d>1$). From the observability of the node dynamics, the coupling function 
between the nodes, and the adjacency matrix, it is indeed possible to construct 
the observability of a large network with an arbitrary topology. 

\end{abstract}

\section{Introduction}

Consider a network composed of $N$ nodes each one of them having a 
$d$-dimensional dynamics and whose interactions are given by an adjacency 
matrix $A$. We can thus distinguish three levels of description of this 
network: i) the node dynamics using the corresponding $d \times d$ node 
Jacobian matrix ${\cal J}_{\rm n}$, ii) the topology described by the $N \times 
N$ adjacency matrix $A$, and iii) the whole dynamical network described by the  $d \cdot N \times d \cdot N$ network Jacobian matrix ${\cal J}_{\rm N}$. There 
are two possible conventions for writing the adjacency matrix, one being the 
transposed of the other. In order to do this without unnecessary complicated 
notations, we will retain the convention used by Newman \cite{New10} in which 
each element $A_{ij}$ of the adjacency matrix $A$ corresponds to an edge from 
node $j$ to node $i$.

The node Jacobian matrix ${\cal J}_{\rm n}$,  computed from the set of the $d$ 
differential equations governing the node dynamics, allows an easy construction 
of the fluence graph describing how the $d$ variables of the node dynamics are 
interacting. Such fluence graphs were used 
by Lin for assessing 
the controllability of linear systems \cite{Lin74} and later on the theory was 
extended to address their observability \cite{Cha92}. When dealing with 
dynamical networks it is important to distinguish the adjacency matrix $A$ from 
the network Jacobian matrix ${\cal J}_{\rm N}$ since, very often the 
observability of a network has been wrongly investigated by only taking into 
account the adjacency matrix \cite{Bia17,Has13,Mie09} and disregarding the node 
dynamics. We 
show how such an approach does not always provide correct results.

Without loss of generality, we will exemplify our methodology to assess the 
observability of dynamical networks by considering networks of diffusively 
coupled R\"ossler systems \cite{Ros76c} ($d=3$). The knowledge gathered from 
the analysis of dyads and triads of R\"osslers will guide us to propose some 
rules to handle larger networks in a systematic way by decomposing the networks 
in blocks whose observability properties is known.
In order to select a reduced set of variables we will use a graphical approach
by introducing a pruned fluence graph of the network Jacobian matrix ${\cal J}_{\rm N}$ as developed in \cite{Let18b}. Then, the symbolic observability
coefficients are computed as detailed in \cite{Let18} and, when
full observability is detected, the analytical determinant of the observability 
matrix is checked to rigorously validate the graphical and symbolic results.

\section{Theoretical background}

\subsection{Observability matrix}

Let us consider a $d \cdot N$-dimensional network ${\cal N}$ composed of $N$ 
nodes each one having an associated $d$-dimensional dynamics. The network state 
is represented by the state vector $\mb{x} \in \bbR^{d \cdot N}$ whose 
components are given by
\begin{equation}
  \label{system}
  \dot{x}_i = f_i (x_1,x_2,x_3, ..., x_{d \cdot N}) , ~~~~~ (i=1,2,...,d \cdot N)
\end{equation}
where $f_i$ is the $i$th component of the vector field $\mb{f}$. The 
corresponding network Jacobian matrix 
$J_{ij}=\frac{\partial f_i}{\partial x_j}$, can be expressed as
\begin{equation}
  \label{jacnet}
  {\cal J}_{\rm N}= {\mathbb{I}}_N \otimes {\cal J}_n - \rho( L \otimes H)
\end{equation}
reflecting its structure in $N$ diagonal blocks containing the node Jacobian 
matrix ${\cal J}_{\rm n}$. The second term corresponds to the contribution to 
the network dynamics from the topology encoded in the Laplacian matrix 
$ L=(L_{ij})=(A_{ij}-k_i\delta_{ij})$ and the linear coupling function 
$H \in \mathbb{R}^{d \times d}$. $\mathbb{I}_N$ is the square identity matrix 
of size $N$, the symbol $\otimes$ stands for the Kronecker product and $\rho$ 
is the coupling constant. 

Let us introduce the measurement vector $h(\mb{x}) \in \mathbb{R}^m$ whose $m$ 
components are the measured variables. The observability cannot be stated only
from these $m$ measured variables. Indeed, to construct the observability 
matrix ${\cal O}$ of the network dynamics from these $m$ measured variables, it
is also necessary to specify the $d_{\rm r} -m$ variables required for 
completing the vector $\mb{X} \in \mathbb{R}^{d_{\rm r}}$ spanning the 
reconstructed space in which the dynamics is investigated. For these reasons, 
and as introduced by Lin \cite{Lin74}, we will speak about the observability of 
the pair $[{\cal J}_{\rm N}, \mb{X}]$ to explicit the fact that the network 
described by ${\cal J}_{\rm N}$ is observable {\it via} the $m$ measured 
variables and $d_{\rm r} -m$ of their derivatives. Since we are interested in 
the smallest state space in which the dynamics can be investigated, we 
will limit ourselves to the case where $d_{\rm r} = d \cdot N$. 

The observability of a dynamical network can be defined as follows.
Let us consider the case when $m=1$ (a generalization to larger 
$m$ is straightforward), and let $\mb{X} \in \mathbb{R}^{d\cdot N}$ be the 
vector spanning the reconstructed space  obtained by using the $(d\cdot N-1)$ 
successive 
Lie derivatives of the measured variables. The dynamical system (\ref{system}) 
is said to be {\it state observable} at time $t_f$ if every initial state 
$\mb{x} (0)$ can be uniquely determined from the knowledge of a finite time 
series $\{ \mb{X} \}_{\tau = 0}^{t_f}$. In practice, it is possible to test 
whether the pair $[ {\cal J}_{\rm N}, \mb{X} ]$ is observable by computing the 
rank of the observability matrix \cite{Her77}, that is, the Jacobian matrix of 
the Lie derivatives of $h(\mb{x})$. 

Differentiating the measured vector $h(\mb{x})$ yields 
$\frac{{\rm d}}{{\rm d}t}h(\mb x) 
   =\frac{\partial h}{\partial \mb{x}}\mb{f}(\mb{x})
   ={\cal L}_{\mb f}\, h(\mb{x}) $,
where ${\cal L}_{\mb{f}}\, h(\mb x)$ is the Lie derivative of $h(\mb{x})$ along 
the vector field $\mb f$. The $k$th order Lie derivative is given by
%
${\cal L}_{\mb{f}}^k h(\mb x)=\frac{\partial {\cal L}_{\mb{f}}^{k-1}h(\mb
  x)}{\partial \mb x}{\mb f}(\mb x)$,
${\cal L}_{\mb f}^0\, h(\mb x)=h(\mb x)$ being the zeroth order Lie derivative 
of the measured variable itself. Therefore, the $d \cdot N \times d \cdot N$ 
observability matrix ${\cal O}_{\mb X}$ can be written as
%
%
\begin{equation}
  {\cal O}_{\mb X} (\mb x) = 
  \left[
      {\rm d} h(\mb{x}), {\rm d} {\cal L}_{\mb{f}} \, h(\mb{x}) , \dots, {\rm d} {\cal L}^{d-1}_{\mb{f}} h(\mb{x}) 
  \right]^T \, . 
\end{equation}
The pair $[ {\cal J}_{\rm N}, \mb{X} ]$ is 
state observable if and only if  the observability matrix has full rank, that 
is, rank$({\cal O}_{\mb X})=d \cdot N$. The Jacobian matrix of the coordinate 
transformation $\Phi: \mathbb{R}^{d \cdot N} \mapsto \mathbb{R}^{d\cdot N}$ 
between the original state space $\mathbb{R}^{d\cdot N} (\mb{x})$ and the 
reconstructed space $\mathbb{R}^{d\cdot N} (\mb{X})$ is 
the observability matrix ${\cal O}_{\mb X}$ \cite{Let05a}. 

\subsection{Symbolic observability coefficients}

The procedure to calculate the symbolic observability coefficients is
implemented in four steps as follows \cite{Let18}. The first step is devoted 
to the construction of the symbolic Jacobian matrix 
$\tilde{\cal J}_{\rm N}$ by replacing each constant element $J_{ij}$ by ``1'', 
each non-constant polynomial element $J_{ij}$ by ``$\1$'', and each rational 
element $J_{ij}$ by ``$\2$'' when the $j$th variable is present in the 
denominator or by $\1$ otherwise. Rational terms are distinguished from 
non-constant polynomial terms since they strongly reduce the observability.

The second step corresponds to the construction of the symbolic observability 
matrix $\tilde{\cal O}_{\mb X}$ \cite{Let18}. 
When $m$ variables are measured, the construction of $\tilde{\cal O}_{\mb X}$ 
is performed by blocks of size $(\kappa_i + 1)\times d$, being $\kappa_i$ the 
number of derivatives of the $i$th measured variable and 
$m + \sum_{i=1}^m \kappa_{i} = d \cdot N$: the construction of each block 
follows the same rules as introduced in \cite{Let18} for univariate 
measures. 

The third step consists in computing the symbolic observability coefficients. 
The determinant of $\tilde{\cal O}_{\mb X}$ is computed according to the 
symbolic algebra defined in \cite{Let18} and expressed as products and 
addends of the symbolic terms $1$, $\1$ and $\2$, whose number of occurrences 
are stored in variables $N_1$, $N_{\1}$ and $N_{\2}$, respectively. A special
condition is required for rational systems such that, if $N_{\1}=0$ and
$N_{\2}\neq 0 $ then $N_{\1}=N_{\2}$. The symbolic observability coefficient
for the  reconstructed vector ${\mb X}$ is then equal to
%
$   \eta_{\mb X} = \displaystyle \frac{1}{D}{N_1} +\frac{1}{D^2}N_{\1}+\frac{1}{D^3} N_{\2}  $
with $D=\mbox{max }(1,N_1) + N_{\1} + N_{\2}$ and $0\le\eta_{\mb X}\le 1$,
being $\eta_{\mb X}=1$ for a reconstructed vector $\mb{X}$ providing a full 
observability. It was shown that the observability can be considered as being
good when $\eta_{\mb X} \geq 0.75$ \cite{Sen16}.

\subsection{Selecting the variables to measure}

A systematic check of all the possible combinations of $m$ measured variables
and their $d\cdot N-m$ derivatives turns out to be a daunting task for large 
$N$ and large $d$. Therefore, it becomes crucial to furnish methods to unveil a 
tractable set 
of variables providing full observability of a system. This may be achieved by
using a graphical approach \cite{Let18b} which is an improved version of the 
procedure introduced by Liu {\it et al.} \cite{Liu13}. A pruned fluence graph 
with $d \cdot N$ vertices (one per variable) and a directed edge 
$x_j\rightarrow x_i$ is drawn between variables $x_j$ and $x_i$ when the element 
$J_{ij}$ of ${\cal J}_{\rm N}$ is constant. At 
least one variable from each root Strongly Connected Component (rSCC) of the 
pruned fluence graph has to be measured \cite{Let18b}. A rSCC is a subgraph in 
which there is a directed path from each node to every other node in the subgraph and 
with no {\it outgoing} edges. As we will see, a pruned fluence graph provides 
a necessary but not a sufficient reduced set of variables to measure for 
getting an observable pair $[ {\cal J}_{\rm N}, \mb{X} ]$.

\section{Observability of the node dynamics}

\begin{figure}[ht]
  \centering
  \includegraphics[width=0.15\textwidth]{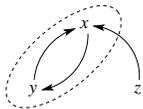} \\[-0.3cm]
  \caption{Pruned fluence graph of the R\"ossler system where an edge is 
drawn between variables $x_i$ and $x_j$ whenever $J_{ij}$ is a nonzero 
constant. A dashed oval surrounds the root strongly connected component 
(rSCC).  Edges $x_i \rightarrow x_i$ are omitted since they do not contribute 
to the determination of the rSCC. }
  \label{pfgros}
\end{figure}

The node dynamics corresponds to the R\"ossler system \cite{Ros76c} $(x_1,x_2,x_3)=(x,y,z)$ whose evolution is governed by the vector field $(f_1,f_2,f_3)=[-y-z,x+ay,b+z(x-c)]$, 
 whose Jacobian matrix is 
\begin{equation}
  {\cal J}_{\rm n} = 
  \left[
    \begin{array}{ccc}
      0 & -1 & -1 \\[0.1cm]
      1 & a & 0 \\[0.1cm]
      z & 0 & x-c 
    \end{array}
  \right] \, .
\end{equation}
Its nonzero constant elements $J_{ij}$ lead to the pruned fluence graph shown 
in Fig.\ \ref{pfgros} which has a single rSCC (dashed oval) containing variables $x$
and $y$.

Variable $z$ can thus be discarded from measurements but, at least, variable 
$x$ or $y$ must be measured. The symbolic observability coefficients for the 
pair $[ {\cal J}_{\rm n}, (x, \dot{x}, \ddot{x}) ]$, 
$[ {\cal J}_{\rm n}, (y, \dot{y}, \ddot{y}) ]$, and
$[ {\cal J}_{\rm n}, (z, \dot{z}, \ddot{z}) ]$ are 
$\eta_{x \dot{x} \ddot{x}} = 0.86$, $\eta_{y \dot{y} \ddot{y}} = 1.00$, and
$\eta_{z \dot{z} \ddot{z}} = 0.44$, respectively. 
This means that the pair $[ {\cal J}_{\rm n}, (y, \dot{y}, \ddot{y}) ]$ is 
fully observable, the observability of the pair $[ {\cal J}_{\rm n}, (x, 
\dot{x}, \ddot{x}) ]$ is good and the pair $[ {\cal J}_{\rm n}, (z, \dot{z}, 
\ddot{z}) ]$ is poorly observable. The pruned fluence graph returns the two 
variables providing the largest observability coefficients when each one of 
them is measured alone. This can be analytically confirmed by computing the 
determinants of the corresponding observability matrices which are 
Det~${\cal O}_{x \dot{x} \ddot{x}} = x-(a+c)$, 
Det~${\cal O}_{y \dot{y} \ddot{y}} = 1$, and
Det~${\cal O}_{z \dot{z} \ddot{z}} = z^2$, respectively. If 
Det~${\cal O}_{\mb X} = 0$ 
for a subset ${\cal M}^{\rm obs} \subset \mathbb{R}^d$ of the state space
associated with the node dynamics, then ${\cal M}^{\rm obs}$ is non observable 
through the measurements and it is called the 
singular observability manifold.  Since Det~${\cal O}_{y \dot{y} \ddot{y}} 
= 1$, ${\cal M}^{\rm obs}$ is an empty set, the pair 
$[ {\cal J}_{\rm n}, (y, \dot{y}, \ddot{y}) ]$ is actually fully observable. 
When the reconstructed space is spanned by $\mb{X} = (x, \dot{x}, \ddot{x})$, 
the plane defined by $x = a+c$ is non observable. The plane $z=0$ is 
nonobservable when $z$ is measured. It was shown that the complexity of the 
determinant, assessed for instance by the order of its expression (1 for
Det~${\cal O}_{x \dot{x} \ddot{x}}$, 0 for Det~${\cal O}_{y \dot{y} \ddot{y}}$, 
and 2 for Det~${\cal O}_{z \dot{z} \ddot{z}}$), is related to the observability:
the larger the order, the less observable the pair $[ {\cal J}_{\rm n},
\mb{X} ]$ \cite{Let02}. 

\section{Observability of small network motifs}

\subsection{Dyads ($N=2$)}

Let us start with a small network motif of two R\"osslers bidirectionally coupled by either $x$, $y$, or $z$. From the analysis of this basic motif we will derive 
general rules for assessing the observability of larger networks. The corresponding ${\cal J}_{\rm N}$ for the case the two nodes are coupled through the $x$ variable is given by

\[
  {\cal J}_{\rm N}= 
  \left[
    \begin{array}{ccc|ccc}
      -\rho_x & -1 & -1 & \rho_x & 0 & 0\\[0.1cm]
      1 & a & 0 & 0 & 0 & 0 \\
      z & 0 & x-c & 0 & 0 & 0\\\hline
      \rho_x & 0 & 0 & -\rho_x & -1 & -1\\
      0 & 0 & 0 & 1 & a & 0\\
      0 & 0 & 0 & z & 0 & x-c\\      
    \end{array}
  \right] \, 
\]
where $H=H_x=[1,0,0;0,0,0;0,0,0]$ has been used in  Eq.~(\ref{jacnet}).

\begin{figure}[ht]
  \centering
  \begin{tabular}{ccccc}
    \includegraphics[width=0.24\textwidth]{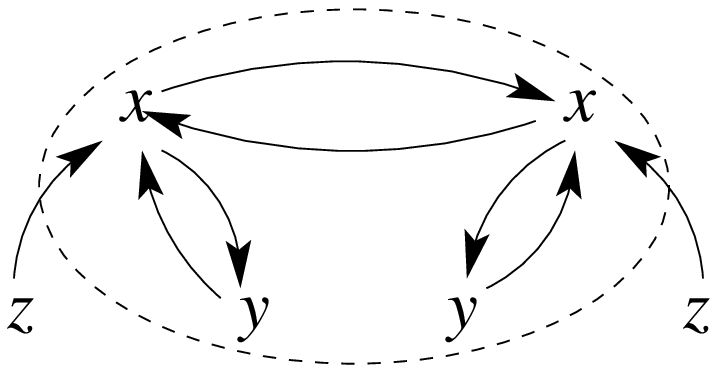} & ~~~~ &
    \includegraphics[width=0.22\textwidth]{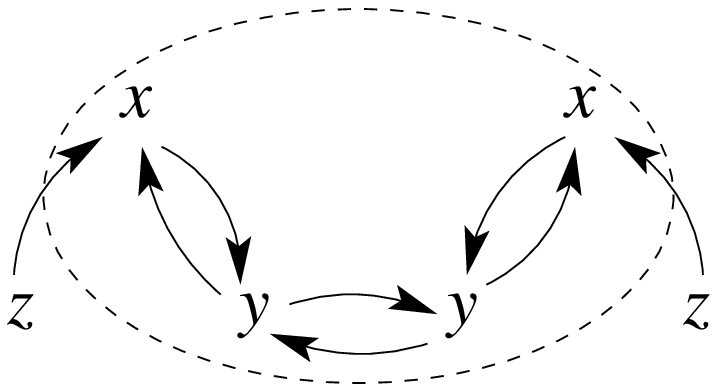} & ~~~~ &
    \includegraphics[width=0.29\textwidth]{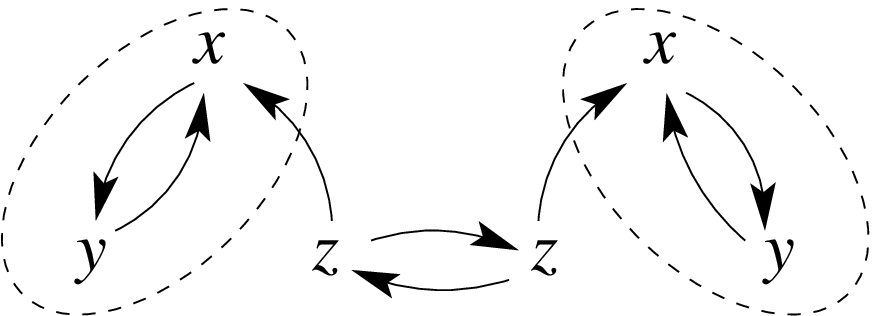} \\
    \includegraphics[width=0.19\textwidth]{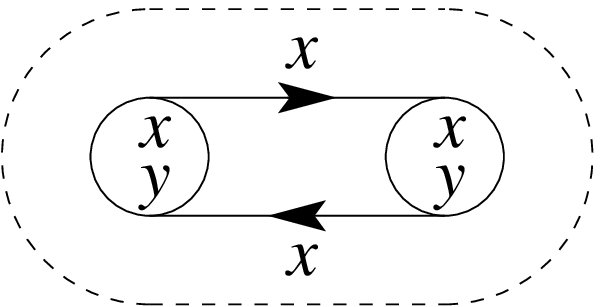} &~~ &
    \includegraphics[width=0.19\textwidth]{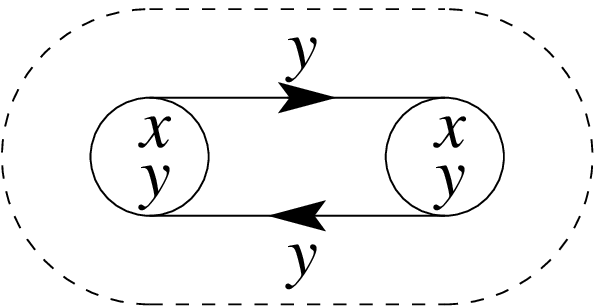} &~~ &
    \includegraphics[width=0.17\textwidth]{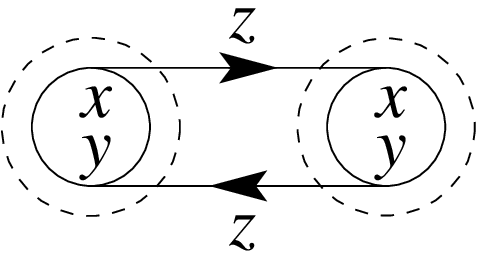} \\
    (a) $\rho_x \neq 0$: 1 rSCC &~~ &
    (b) $\rho_y \neq 0$: 1 rSCC &~~ &
    (c) $\rho_z \neq 0$: 2 rSCC \\[-0.2cm]
  \end{tabular}
  \caption{Pruned fluence graphs (top) and network connection motifs (bottom)
for small networks motifs ($N=2$) of R\"ossler systems coupled by their 
different variables. The root strongly connected components (rSCC) are 
shown in dashed lines.}
  \label{ros2}
\end{figure}

Figure \ref{ros2} shows the pruned fluence graphs obtained from ${\cal J}_{\rm N}$ for the three coupling configurations. Below each graph, a compact representation at the level of the adjacency matrix is also provided, indicating as well the coupling nature of the bidirectional links. 
There is only one rSCC when the two nodes are coupled either {\it via} variable 
$x$ or $y$ whereas there are two rSCCs when coupled {\it via} variable $z$. 
This suggests that at least one variable has to be measured among 
$\{ x_1, y_1, x_2, y_2 \}$ in the first two cases (Figs. \ref{ros2}a and 
\ref{ros2}b) and one ($x_i$ or $y_i$) in each rSCC in the latter case. In the 
following, we will analyze in detail the cases where $m=1,2$, that is, 1 or 2 
measured variables, which can be acquired in $N_m=1$ or 2 nodes. We computed 
the symbolic observability coefficients $\eta_{\mb X}$ for all possible 
reconstructed vectors $\mb{X}$. A summary of these analysis is reported in 
Table \ref{resros2}.

When a single variable is measured ($m=1$), the pair $[{\cal J}_{\rm N}, \mb{X} ]$ is 
always poorly observable, even when there is a single rSCC (via $H_x$ or $H_y$).
The symbolic observability coefficients are very well confirmed by the determinants which are at 
least second-order polynomials (not shown).

When two variables are measured, $m=2$, in just a single node, $N_m=1$, full observability of the pair 
$[{\cal J}_{\rm N}, \mb{X}]$ is obtained only through $H_y$. We found three 
possibilities for the reconstructed state vector $\mb{X}$ providing $\eta_{\mb{X}}=1$.
In these cases, the corresponding determinants depend on $\rho_y^3$ (Table.\ \ref{resros2}). Such a strong dependency on the coupling strength could deteriorate the observability 
when $\rho_y$ becomes small. On the other hand, when the two variables measured are coming from two different nodes, there is a wide variety of possibilities providing a fully 
observable pair $[{\cal J}_{\rm N}, \mb{X} ]$.  There is a 
strong advantage of using variable $y$ and its first two derivatives in each node 
since it provides a full observability and the determinant is not dependent on 
the coupling strength (Det~${\cal O}_{\mb X} = 1$). 

\begin{table*}[ht]
  \centering
  \caption{Symbolic observability coefficients $\eta$ for the dyads shown in Fig.~\ref{ros2}.
  The type of coupling function $H$, the number $m$ of measured variables and the number of nodes $N_m$ where they are measured are also 
reported. Analytical determinants are reported only in those cases when $\eta = 1$. To shorten the 
notation of the reconstructed vector, we used $y^3$ instead of $(y, \dot{y}, \ddot{y})$, where the exponent refers to the number of derivatives (including the variable itself). The index is omitted when only the variable itself  appears in the reconstruction vector.}
\label{resros2}
  \begin{tabular}{lllrlr}
    \\[-0.5cm]
    \hline \hline
    \\[-0.3cm]
    $H$ \hspace*{0.4cm}& 
    $m=1$ & 
    \multicolumn{2}{c}{$m=2$, $N_m = 1$} & 
    \multicolumn{2}{c}{$m=2$, $N_m = 2$} \\[0.1cm]
    \hline
    \\[-0.4cm]
    &
    $\eta_{x^6} = 0.65$\hspace*{0.2cm} &  
    $\eta_{y^5z} = 0.91$  &  &  
   \hspace*{0.2cm} $\eta_{y_1^3y_2^3} = 1$ & (Det~${\cal O} = 1$) \\[0.1cm]
    $H_x$ & 
    $\eta_{y^6} = 0.41$ & & & 
\hspace*{0.2cm}    $\eta_{y_1^2y_2^4} = 1$ & (Det~${\cal O} = \rho_z$) \\[0.1cm]
    &
    $\eta_{z^6} = 0.03$ & & & 
\hspace*{0.2cm}    $\eta_{x_1 y_2^5} = 0.91$ & \\[0.1cm]
    & & & & 
\hspace*{0.2cm}    $\eta_{x_1^3 x_2^3} = 0.79$ & \\[0.2cm]
    \hline
    \\[-0.4cm]
    &
    $\eta_{x^6} = 0.66$ & 
    $\eta_{x^5y} = 1$ & (Det~${\cal O} = -\rho_y^3$) & 
 \hspace*{0.2cm}   $\eta_{y_1^3y_2^3} = 1$ & (Det~${\cal O} = 1$) \\[0.1cm]
    $H_y$ & 
    $\eta_{y^6} = 0.56$ & 
    $\eta_{x^2y^4} = 1$ & (Det~${\cal O} = \rho_y^3$) & 
\hspace*{0.2cm}    $\eta_{y_1^2y_2^4} = 1$ & (Det~${\cal O} = -\rho_y$) \\[0.1cm]
    & 
    $\eta_{z^6} = 0.31$ & 
    $\eta_{x^5z} = 1$ & (Det~${\cal O} = \rho_y^3$) &  
   \hspace*{0.2cm} $\eta_{x_1^2y_2^4} = 1$ & (Det~${\cal O} = -\rho_y^3$) \\[0.1cm]
    &  &
    $\eta_{y^4z^2} = 0.86$ & &
\hspace*{0.2cm}    $\eta_{x_1^3y_2^3} = 0.91$ \\[0.1cm]
    &  &
    $\eta_{y^5z} = 0.77$ & &
\hspace*{0.2cm}    $\eta_{x_1^3x_2^3} = 0.91$ \\[0.2cm]
    \hline
    \\[-0.4cm]
    &
    $\eta_{x^6} = 0.72$ & 
    $\eta_{x^5y} = 0.72$ & &  
    \hspace*{0.2cm}$\eta_{y_1^3y_2^3} = 1$ & (Det~${\cal O} = 1$) \\[0.1cm]
    $H_z$ & 
    $\eta_{y^6} = 0.37$ & 
    $\eta_{x^5z} = 0.72$ & &
\hspace*{0.2cm}    $\eta_{y_1^2y_2^4} = 1$ & (Det~${\cal O} = -\rho_z$) \\[0.1cm]
    & 
    $\eta_{z^6} = 0.11$ & 
    $\eta_{x^2z^4} = 0.72$ & &
\hspace*{0.2cm}    $\eta_{x_1^2y_2^4} = 1$ & (Det~${\cal O} = -\rho_z$) \\[0.1cm]
    & &
    $\eta_{y^2z^4} = 0.72$ & &  
\hspace*{0.2cm}    $\eta_{y_1 y_2^5} = 0.86$ \\[0.1cm]
    &
    & & & 
\hspace*{0.2cm}    $\eta_{x_1^3 x_2^3} = 0.79$ \\[0.2cm]
    \hline \hline
  \end{tabular}
\end{table*}

Among other possibilities offering full observability is that from the reconstructed vector
$\mb{X} = (x_1^2 \, y_2^4)$ (where the notation $x_i^j$ designates the 
first $j$ Lie derivatives of variable $x_i$, being the first one the variable 
itself), either using the coupling functions $H_y$ or $H_z$. However, when looking at the corresponding determinants $\Delta_{x_1^2 y_2^4} = \rho_y^3$ and $\Delta_{x_1^2 y_2^4} = - \rho_z$ respectively, 
 we unexpectedly notice that the observability depends on the 
coupling strength in a weaker way when nodes are coupled {\it via} variable $z$ 
than {\it via} variable $y$. And even more surprising is the case when nodes are coupled {\it via} variable $x$, since the determinant (not shown in Table ~\ref{resros2}) $\Delta_{x_1^2 y_2^4} = 0$, indicating that the network is not observable at all. Consequently, the observability of our network with a given topology and reconstructed state vector $\mb{X}$ strongly depends on the coupling variable.

Notice also that the graphical analysis of the pruned fluence graph of 
${\cal J}_{\rm N}$ is providing a necessary condition about the variables to be 
measured for getting full observability but it may not be sufficient. For 
example, in Fig.~\ref{ros2}b, it is recommending to measure either $x$ or $y$ 
from the single rSCC but to get full observability a second variable is needed. 
This leads 
to 
the following propositions.

\begin{prop}
  \label{propmin}
The minimal number $m_{\rm min}$ of variables necessary to measure for getting
full observability of a $d \cdot N$-dimensional network ${\cal N}$ is equal to 
the number $N_{\rm r}$  of root strongly connected components. 
Each measured variable has to be chosen in a different root strongly connected component.
\end{prop}

\begin{coro}
If additional variables are required to get a full observability of a $d \cdot 
N$-dimensional network, they will be selected in the $N_{\rm r}$ root strongly 
connected components and, preferably, in those whose cardinality is the largest. 
\end{coro}

Thus, with these rules from the analysis of the pruned fluence graph,
the number of vectors $\mb{X}$ is sufficiently reduced to make exhaustive computations of the 
symbolic observability coefficients. 

\begin{prop}
  \label{propfully}
When the node dynamics is fully observable from one of its variables, then the 
network ${\cal N}$ is fully observable if that variable is measured at each 
node ($m=N$), independently from the coupling function and topology, even when 
the network is not completely connected. 

\end{prop}


\begin{coro}
When the number $N_m$ of measured nodes is such that $N_m  < N$, by definition,
the choice of the variables to measure is not only dependent on the adjacency 
matrix $A$ and coupling function $H$ but also on the node dynamics.
\end{coro}

\begin{prop}
  \label{propz}
When a network ${\cal N}$ of R\"ossler systems coupled by the variable $z$, 
then $m=N$ nodes must be necessarily measured for getting full observability.
\end{prop}

\subsection{Triads ($N=3$)}

\begin{figure}[ht]
  \centering
  \begin{tabular}{cccccccc}
    \includegraphics[width=0.11\textwidth]{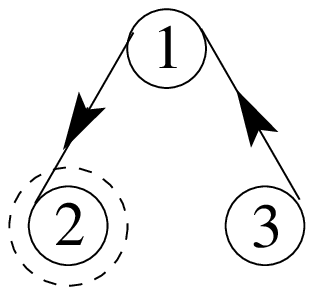} & 
    \includegraphics[width=0.11\textwidth]{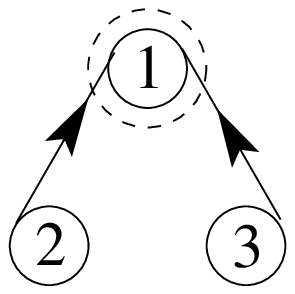} &
    \includegraphics[width=0.11\textwidth]{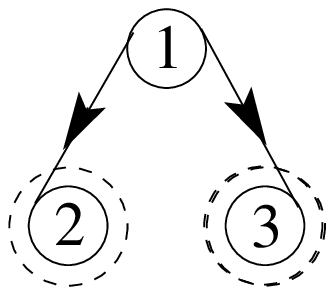} &
    \includegraphics[width=0.11\textwidth]{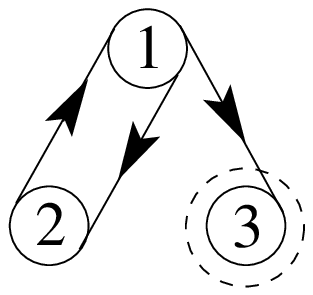} &
    \includegraphics[width=0.11\textwidth]{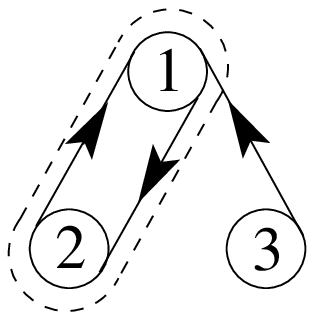} & 
    \includegraphics[width=0.11\textwidth]{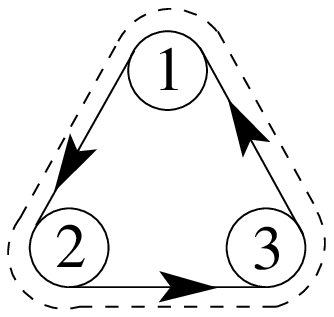} &
    \includegraphics[width=0.11\textwidth]{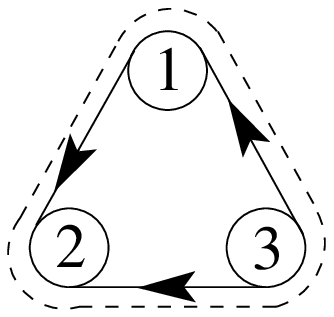} \\
    (a) T$_{\rm 2a}$ & 
    (b) T$_{\rm 2b}$ & 
    (c) T$_{\rm 2c}$ & 
    (d) T$_{\rm 3a}$ & 
    (e) T$_{\rm 3b}$ & 
    (f) T$_{\rm 3c}$ & 
    (g) T$_{\rm 3d}$ \\ 
    \includegraphics[width=0.11\textwidth]{nodros3x3l-4.eps} & 
    \includegraphics[width=0.11\textwidth]{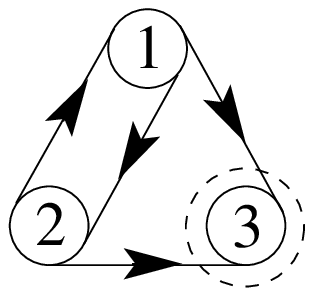} & 
    \includegraphics[width=0.11\textwidth]{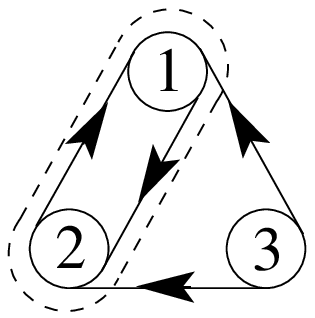} &
    \includegraphics[width=0.11\textwidth]{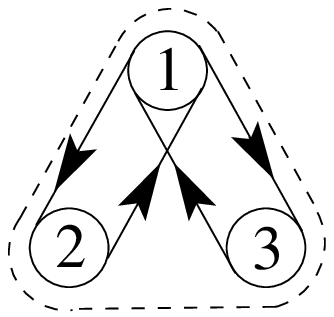} & 
    \includegraphics[width=0.11\textwidth]{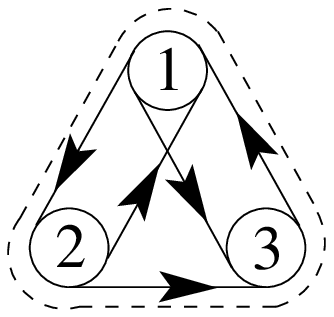} &
    \includegraphics[width=0.12\textwidth]{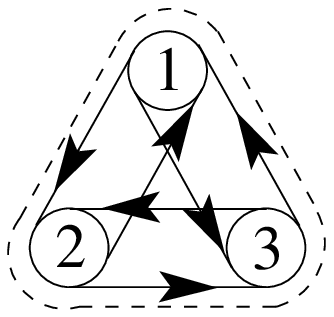} \\
    (h) T$_{\rm 4a}$ & 
    (i) T$_{\rm 4b}$ & 
    (j) T$_{\rm 4c}$ & 
    (k) T$_{\rm 4d}$ & 
    (l) T$_{\rm 5}$ & 
    (m) T$_{\rm 6}$ \\[-0.2cm] 
  \end{tabular}
  \caption{Network connection motifs for triad networks ($N=3$) of R\"ossler 
systems coupled by variable $x$ or $y$. Only the rSCCs are shown (dashed 
line).} 
  \label{ros3}
\end{figure}

Let us now consider motifs of $N=3$ nodes. To limit the number of cases to discuss, we will analyze only
triad networks coupled through variable $y$ since this is the sole coupling configuration for which a dyad of R\"osslers are fully observable from measurements in a single node (see Table \ref{resros2}). In order to refer to all the possible triad motifs shown in Fig.~\ref{ros3}, we will distinguish them by the number $l$ of directed edges, T$_{\rm l}$,  such that there are five classes of motifs: T$_{\rm 2}$ (3), T$_{\rm 3}$ (4), T$_{\rm 4}$ (4), T$_{\rm 5}$ (1), and T$_{\rm 6}$ (1).


Let us start with the triad T$_{\rm 2a}$  shown in Fig.\ \ref{ros3}a. There is 
a single root strongly connected component comprised by the vertices $x_2$ and 
$y_2$. According to this graph, measuring either $x_2$ or $y_2$ in node 2 
should provide full observability of the triad T$_{\rm 2a}$. However, when two 
variables are measured in that node, the largest observability coefficient is 
$\eta_{y_2^8 z_2} = 0.59$. Measuring a third variable in node 2 does not 
improve the observability since the symbolic observability coefficient becomes 
null. The triad T$_{\rm 2a}$ is therefore poorly observable when measurements 
are only performed in the rSCC. Therefore, a 
proposition is made as follows.

\begin{prop}
  \label{prop1}
In a network 
of $N$ R\"ossler systems, it is not possible to reconstruct with full 
observability the space associated with three nodes from measurements in a 
single node. 
\end{prop}

\begin{table}[ht]
  \centering 
  \caption{Determinants of the observability matrix and symbolic observability 
coefficients for three of the triads shown in Fig.~\ref{ros3} and for different reconstructed vectors
$\mb{X}$. Determinants with a polynomial of degree $i$ dependence are indicated with $P_i$. The value of the observability coefficient with a $^*$ is spurious
due to symmetries in the observability matrix that cancel the determinant and
that the symbolic formalism does not detect: it should be zero.}
  \label{restri}
  \begin{tabular}{ccccccc}
    \\[-0.3cm]
    \hline \hline
    \\[-0.3cm]
    Triad & & T$_{\rm 2a}$ & & T$_{\rm 4d}$ & & T$_6$ \\[0.1cm]
    \hline 
    \\[-0.3cm]
    $\mb{X} = (x_2^2, y_2^4, y_3^3)$ & ~~~ &
    Det~${\cal O}_{\mb X} = -\rho_y^3$ &~~~ &
    Det~${\cal O}_{\mb X} = -\rho_y^3$ &~~~ &
    Det~${\cal O}_{\mb X} = -\rho_y^3$ \\
    & & $\eta_{\mb X} = 1$ & & 
    $\eta_{\mb X} = 1$ & & 
    $\eta_{\mb X} = 1$ \\[0.1cm]
    $\mb{X} = (x_2^2, y_2^5, y_3^2)$ & ~~ &
    Det~${\cal O}_{\mb X} = -\rho_y^5$ &~~ &
    Det~${\cal O}_{\mb X} = -\rho_y^5$ &~~ &
    Det~${\cal O}_{\mb X} = \rho_y^4 P_1$ \\
    & & $\eta_{\mb X} = 1$ & & 
    $\eta_{\mb X} = 1$ & & 
    $\eta_{\mb X} = 0.88$ \\[0.1cm]
    $\mb{X} = (x_2^2, y_2^6, y_3)$ & ~~ &
    Det~${\cal O}_{\mb X} = \rho_y^7 P_2$ &~~ &
    Det~${\cal O}_{\mb X} = \rho_y^7 P_3$ &~~ &
    Det~${\cal O}_{\mb X} = \rho_y^7 P_3$ \\
    & & $\eta_{\mb X} = 0.82$ & & 
    $\eta_{\mb X} = 0.82$ & & 
    $\eta_{\mb X} = 0.68$ \\[0.1cm]
    $\mb{X} = (x_2^2, y_2^7)$ & ~~ &
    Det~${\cal O}_{\mb X} = 0$ &~~ &
    Det~${\cal O}_{\mb X} = \rho_y^9 P_5$ &~~ &
    Det~${\cal O}_{\mb X} = \rho_y^7 P_5$ \\
    & & $\eta_{\mb X} = 0.69^*$ & & 
    $\eta_{\mb X} = 0.69$ & & 
    $\eta_{\mb X} = 0.60$ \\[0.1cm]
    \hline \hline
  \end{tabular}
\end{table}

Therefore, the nine dimensions of a R\"ossler triad can not be observed from just measuring in one node. However, from the dyad analysis, when the coupling function is via the $y$ variable, it is possible to reconstruct the six associated dimensions from measurements ($m=2$) in a single node. In order to investigate what is the largest dimension that can be reconstructed, we consider the vector $\mb{X} = (x_2^2, y_2^4, y_3^3)$ which provides full 
observability of the triad $T_{\rm 2a}$ (Det~${\cal O}_{\mb X} = - \rho_y^3$) by performing $m=3$ measurements, two in node 2 and one in node 3. Now, we proceed by progressively adding an extra Lie derivative of 
$y_2$ and removing it from $y_3$ until full observability is lost for $\mb{X}=(x_2^2,y_2^7)$. For the case  $\mb{X} = (x_2^2, y_2^5, y_3^2)$, Det~${\cal O}_{\mb X} = - \rho_y^5$ and therefore a full observable pair $[ {\cal J}_{\rm N}, \mb{X}]$ is still obtained. However, one more Lie derivative of $y_2$, $\mb{X} = (x_2^2, y_2^6, y_3)$,  leads to
$ \mbox{Det } {\cal O}_{\mb X} = - \rho_y^3
  \left[ \left( \displaystyle a+c -x_1 \right) \, 
         \left( \displaystyle x_1 - x_3 \right) + y_1 + 2z_1 - z_3 -1 
  \right]$, 
that is, observability is good since a singular observability manifold appears 
with this first-order determinant and $\eta_{\mb X} = 0.82$. Therefore, the 
largest dimension that can be reconstructed from measurements in a single 
R\"ossler node is six. The triads T$_{\rm 4d}$ and T$_6$ led to 
similar results (Table\ \ref{restri}): 
full observability of a triad of R\"osslers is only possible when no more than 
seven dimensions are reconstructed from measurements in a single node. As soon 
as eight dimensions are 
recovered from one node, the 
reconstructed vector provides poor observability of the whole system. 

\begin{prop}
  \label{propdiad}
In a network 
of R\"ossler systems, it is not possible to reconstruct with full 
observability more than two nodes from measurements in a single node. 
\end{prop}



\begin{prop}
  \label{prophalf}
When $N>2$ R\"ossler systems are coupled, full observability is only possible 
if at least $N_m = \frac{N}{2} + (N\bmod{2})$ nodes are measured and $m=N$ 
variables are measured.
\end{prop}

This proposition could be specific to the R\"ossler system or even be more 
generic. This will be further investigated elsewhere. 

An additional question to address to complete the observability analysis of the 
triad is to check whether it is possible to reconstruct a node from another one 
not directly connected to it. Let us consider the triad T$_{\rm 2a}$ and the 
reconstructed vector $\mb{X} = (x_2^2, y_2^4, y_1^3)$ where $m=3$ measurements 
are performed in nodes 1 and 2. Note that in triad T$_{\rm 2a}$ information is 
flowing from 3 to 2 through node 1 which is measured. The three extra dimensions
reconstructed from node 2 cannot be used for node 3 (not directly connected)
but only for node 1. Node 1 is thus observed twice, leading to 
Det~${\cal O}_{\mb X} = 0$: there is null observability of the triad 
T$_{\rm 2a}$ from such a reconstructed vector $\mb X$. Contrary to this, 
the vector $\mb{X} = (y_2^3, x_1^2, y_1^4)$ provides full observability of 
triad T$_{\rm 2a}$, node 3 being reconstructed from node 1. Then, we state the 
following proposition.

\begin{prop}
  A necessary condition for having full observability of a non measured node n$_i$ from a measured one n$_j$ is that there is an edge from n$_i$ to n$_j$. 
\end{prop}

\begin{coro}
If the node dynamics is a R\"ossler system, a non measured node can be fully 
observable if it is directly coupled {\rm via} variable $y$ to a measured one.
\end{coro}

\section{Larger networks}

\subsection{Star network}

Let us consider a star network of $N$ nodes, being $N-1$ of them leaves and one 
acting as the hub. The number $N_{\rm r}$ of rSCCs depends on the number $l$ of 
edges and how these edges are directed. When couplings are bidirectional, there 
is a single rSCC that contains all the nodes. When unidirectional couplings are 
all directed to the hub, the hub is the rSCC. When all the edges are out-going 
from the hub, there are $N_{\rm r} = N-1$ rSCC, each one made of one leaf. In a 
random star network with $N_{\rm out}$ edges out-going from the hub, there are 
$N_{\rm out}$ rSCCs, each one made of one of the leaves receiving one of these 
out-going edges. In all cases, according to propositions \ref{propmin} and 
\ref{propdiad}, $m=N$ variables must be measured in $N_m = N_r$ nodes.

\subsection{Ring network}

In a ring network of R\"ossler systems coupled {\it via} variable $y$, 
according to propositions \ref{propmin} and \ref{propdiad}, $m = N$ variables 
must be measured in $N_m = \frac{N}{2} + (N\bmod{2})$ nodes for full 
observability of the pair $[{\cal J}_{\rm N}, \mb{X}]$. This result does not 
depend on the directionality of the edges (they can be either bidirectional or 
unidirectional). Along the ring, one in two nodes are measured. 

\subsection{Random network ($N=28$)}

\begin{figure}[ht]
  \vspace{-0.4cm}
  \centering
  \includegraphics[width=0.45\textwidth]{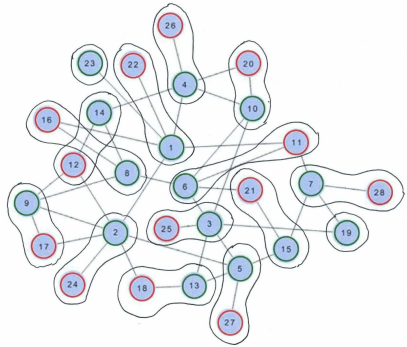} \\[-0.3cm]
  \caption{Topology of the random network ($N=28$) used in Ref.~\cite{Sev16} to 
implement a network of electronic R\"ossler-like circuits. Nodes are grouped by 
pairs 
to get full observability of this network from measurements in $N_m = 15$ 
nodes.}
  \label{topelec}
\end{figure}

We here investigate a random network made of 28 R\"ossler systems 
bidirectionally coupled {\it via} variable $y$ according to the topology 
(Fig.\ \ref{topelec}) of an electronic network 
\cite{Sev16}. Applying propositions
\ref{propmin} and \ref{propdiad}, nodes are grouped by pairs depending on their 
edges. One possibility is to measure $m= N+1$ variables in 15 nodes, namely in 
nodes 1, 2, 3, 4, 5, 
6, 7, 8, 9, 10, 13, 14, 15, 19, and 23. Measurements are needed in $N_m = 
\frac{N}{2} +1$ since two 
nodes, 22 and 23, are only connected to node 1, from which it is not possible
to reconstruct three nodes according to proposition \ref{prop1}. One of these
two nodes must also be  measured. In all nodes but nodes 19 and 23, the 
reconstructed vector is $\mb{X}_i = (x_i^2, y_i^4)$ while in nodes 19 and 23   
$\mb{X}_j = y_j^3$. The symbolic coefficient was 
equal to one and the 
analytical determinant of the observability matrix is 
Det~${\cal O}_{\mb{X}} = \rho_y^{39}$, therefore validating all our results. 
The expression of this determinant could mean that the observability is strongly sensitive to the coupling 
value. Nevertheless, since nodes are grouped by pair, the dependency on the 
coupling value should not be practically worse than the one observed for a 
pair of nodes, that is, depending on $\rho_y^3$.

\section{Conclusion}

We showed that it is possible to construct a procedure to reliably determine
the observability of networks whose node dynamics are structurally identical
(the governing equations have the same functional form but parameter values can 
differ). 
A reduced set of variables to measure in a network of $N$ nodes 
and providing a full observability of it
can be selected using a graphical approach \cite{Let18b}. Then 
symbolic 
coefficients are computed to quantify the observability 
of the network dynamics provided by the measurements \cite{Let18}. To be fully 
reliable, network observability must be investigated from the complete 
Jacobian matrix ${\cal J}_{\rm N}$ of the network which encodes the topology, 
the coupling function and the node dynamics. Nevertheless, 
some systematic rules for assessing the observability of the network can be 
derived   from the node Jacobian matrix ${\cal J}_{\rm n}$ and the coupling 
function of dyads and triads. 
First 
we determine the 
observability of the node dynamics. Then, using the results obtained
from the analysis of a dyad, general rules can be established to be applied 
to larger networks. In the case of R\"ossler systems, 
it is not possible to reconstruct more than two nodes from measurements in one 
node. It is necessary to measure at least in $\frac{N}{2} + (N\bmod{2})$ nodes 
for getting full observability of a network made of $N$-R\"ossler systems 
coupled {\it via} variable $y$. 
For any other coupling, $N$ nodes have to be measured. Therefore, 
the coupling function may critically affect the network observability. 

\section*{Acknowledgments}
ISN acknowledges partial support from the Ministerio de Econom\'ia y Competitividad
of Spain under project FIS2017-84151-P and from the Group of Research 
Excelence URJC-Banco de Santander. 

\bibliographystyle{splncs04}

\bibliography{SysDyn}

\end{document}